\documentclass{nature}

\usepackage{graphicx}

\makeatletter
\let\saved@includegraphics\includegraphics
\AtBeginDocument{\let\includegraphics\saved@includegraphics}
\renewenvironment*{figure}{\@float{figure}}{\end@float}
\makeatother

\usepackage{dcolumn}
\usepackage{bm}
\usepackage[dvipsnames]{xcolor}
\usepackage{float}
\usepackage{amsmath}
\newcommand{\inlineeqnum}{\refstepcounter{equation}~~\mbox{(\theequation)}}
\usepackage{color,soul}

\title{Thermal decoherence and laser cooling of Kerr \\ microresonator solitons}

\author{Tara E. Drake*$^{1,2}$, Jordan R. Stone*$^{1,2}$, Travis C. Briles$^{1,2}$ \& Scott B. Papp$^{1,2,\dagger}$}

\begin{document}
\maketitle
\begin{affiliations}
\item Time and Frequency Division, National Institute of Standards and Technology, Boulder, CO USA
 \item Department of Physics, University of Colorado, Boulder, CO USA
 \\**J.R.S. and T.E.D. share primary authorship of this manuscript.
 \\$^\dagger$Correspondence to: scott.papp@nist.gov
\end{affiliations}

\begin{abstract}
Thermal noise is ubiquitous in microscopic systems and in high-precision measurements. Controlling thermal noise, especially using laser light to apply dissipation, substantially affects science in revealing the quantum regime of gases\cite{anderson1995observation}, in searching for fundamental physics\cite{Parker2018finestructure}, and in realizing practical applications\cite{Rosi2014gravconstant}. Recently, nonlinear light-matter interactions in microresonators have opened up new classes of microscopic devices. A key example is Kerr-microresonator frequency combs\cite{kippenberg2018dissipative}; so-called soliton microcombs not only explore nonlinear science but also enable integrated-photonics devices, such as optical synthesizers\cite{spencer2018optical}, optical clocks\cite{newman2018photonic}, and data-communications systems\cite{marin2017microresonator}. Here, we explore how thermal noise leads to fundamental decoherence of soliton microcombs. We show that a particle-like soliton exists in a state of thermal equilibrium with its silicon-chip-based resonator. Therefore the soliton microcomb's modal linewidth is thermally broadened. Our experiments utilize record sensitivity in carrier-envelope-offset frequency detection in order to uncover this regime of strong thermal-noise correlations. Furthermore, we have discovered that passive laser cooling of the soliton reduces thermal decoherence to far below the ambient-temperature limit. We implement laser cooling by microresonator photothermal forcing, and we observe cooling of the frequency-comb light to an effective temperature of 84 K. Our work illuminates inherent connections between nonlinear photonics, microscopic physical fluctuations, and precision metrology that could be harnessed for innovative devices and methods to manipulate light.

\end{abstract}
\vspace{18pt}
Thermal energy is constantly being exchanged between matter in thermal equilibrium, leading to fluctuations in physical systems. Specifically, in a homogeneous medium at temperature $T$ the thermal fluctuations of an observable $X$ vary by $\left< \delta X^2 \right> = \eta_{X}^2\frac{k_{B}T^{2}}{\rho\, C\, V} \inlineeqnum\label{eq:FD}$, where $\eta_{X}=dX/dT$ is the thermal coupling coefficient, $k_{B}$ is the Boltzmann constant, $\rho$ is the mass density, $C$ is the specific heat, and $V$ the volume\cite{landau1980statistical}. Hence, temperature-dependent observables are stochastic variables with measurement uncertainties imposed by the ambient environment. This fundamental limit has been of interest in low-noise optical metrology in which the measurement sensitivity is set by thermal fluctuations in the optical path length of a cavity or interferometer\cite{numata2004thermal, liu2000thermoelastic}. Thermal physics is well-understood, and progress in mitigating thermal noise has been substantial-- optical resonators and interferometers have facilitated some of the most precise measurements ever made and are directly applied in gravitational-wave detection\cite{abbott2016observation} and optical-atomic timekeeping\cite{ludlow2015optical}. Furthermore, some systems provide a route to lower temperature through laser cooling, whereby random motion is intrinsically damped. Laser cooling revolutionized atomic physics, allowing atoms to be trapped, manipulated, and probed for long periods of time\cite{phillips1998nobel}. In cavity optomechanics\cite{aspelmeyer2014cavity}, photothermal and radiation-pressure forces are used to cool mechanical oscillators for fundamental studies of quantum mechanics\cite{metzger2004cavity, gigan2006self, arcizet2006radiation}. Laser cooling of bulk solids through anti-Stokes fluorescence has also been demonstrated\cite{seletskiy2010laser}.

Monolithic microresonators are an important setting to consider both thermal noise and nonlinear photonics, although the intersection of these regimes remains largely unexplored. Thermo-mechanical (thermal expansion) and thermorefractive (temperature-dependent refractive index) effects contribute to the stochastic fluctuations of microresonator modes\cite{matsko2007whispering}, with microresonator geometry and thermal coefficients defining the practical impact of thermal noise. A seemingly separate consideration is the intensity-dependent index of refraction in microresonators. Nonlinear microresonators have developed largely in parallel to the integrated-photonics revolution\cite{moss2013new}, owing to compatibility with semiconductor processing, and they are increasingly important in classical and quantum optics\cite{strekalov2016nonlinear}. In particular, a Kerr-microresonator excited by a continuous-wave (CW) laser can generate new frequencies or patterns in the intraresonator field. Experiments have harnessed microresonators to generate dissipative Kerr soliton (DKS) frequency combs, which are a stationary nonlinear eigensolution that periodically repeats at the roundtrip time or free-spectral range\cite{kippenberg2018dissipative}. Various aspects of soliton microcombs have been explored, including their threshold properties\cite{Li2018isocontours}, breathing excitations\cite{bao2016observation}, and crystallization\cite{cole2017soliton}. Thermorefractive-noise models can predict phase noise in microcombs, but there have been no quantitative predictions or experiments. 

\begin{figure*}[!t] \centering  \includegraphics[width=\linewidth]{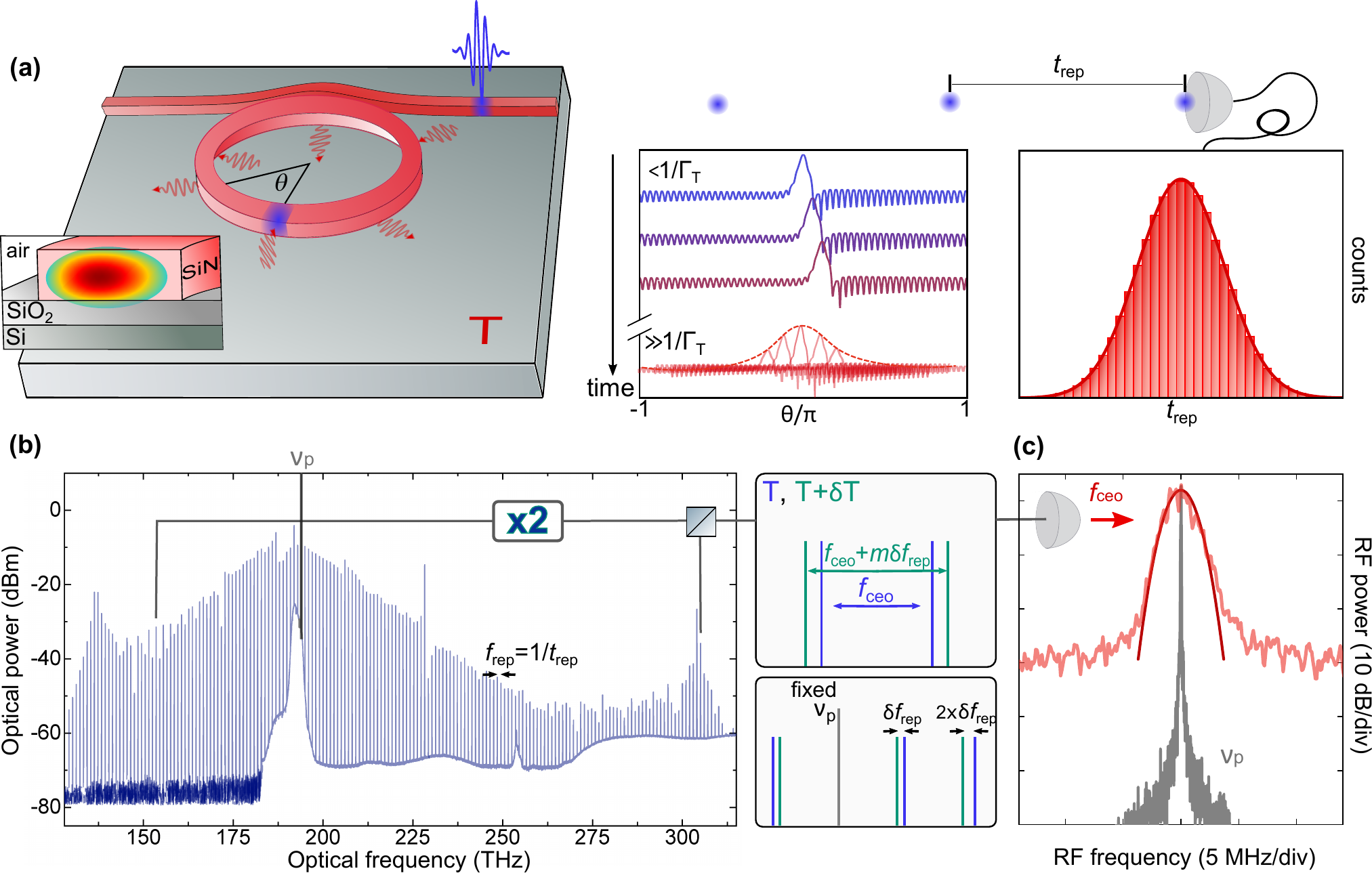}
\caption{Concept of soliton thermal decoherence. (a) Thermal noise couples to a microresonator soliton due to thermorefractive-index fluctuations. The soliton angular position ($\theta$) and the soliton parameters, including $t_{\rm{rep}}$, $f_{\rm{rep}}$, and $f_{\rm{ceo}}$ represented by stochastic ensembles subject to Eq. \ref{eq:FD}, fluctuate after many roundtrips. Emitted soliton pulses fluctuate on the resonator thermal dissipation ($\Gamma_T$) timescale, and measurements of $f_{\rm{ceo}}$ show thermal decoherence. (b) Left panel: Soliton microcomb optical spectrum. Right panels: Fluctuations in $f_{\rm{rep}}$ are amplified in $ f_{\rm{ceo}}$ by the factor $m$, enabling measurements of $f_{\rm{rep}}$. (c) $f_{\rm{ceo}}$ optical lineshape (red data). Our theoretical model (dark red line) of the $f_{\rm{ceo}}$ lineshape at $T=$ 300 K overlays the data. For comparison, we show the optical lineshape of the pump laser (gray).}\label{fig1}
\end{figure*}

Here, we use ultraprecise optical-frequency metrology to provide a clear window into the behavior of Kerr solitons. We report that thermal noise imposes fundamental decoherence on the repetition frequency $(f_{\rm{rep}})$ of a soliton circulating in a microscopic resonator. Therefore, thermal noise defines the measurement uncertainty of the optical-mode frequencies $\nu_n=nf_{\rm{rep}}+f_{\rm{ceo}} \inlineeqnum\label{eq:COMB}$ that comprise the soliton microcomb, where $f_{\rm{ceo}}$ is the carrier-envelope-offset frequency\cite{jones2000carrier} and $n$ is the mode number. Operationally, we use a silicon-nitride (Si$_3$N$_4$, hereafter SiN) ring resonator to create an octave-spanning microcomb, and $f$--$2f$ self-referencing enables precision-enhanced measurements of $n f_{\rm{rep}}$ by the coherence of Eq. \ref{eq:COMB}. These results show that the particle-like soliton-- the frequency-comb itself --is governed by the ambient temperature. Still, the soliton can also be influenced by external fields coupled to the microresonator. We report our observation of laser cooling through passive microresonator forcing, which reduces the soliton microcomb's effective temperature to $84$ K. We apply laser cooling with the photothermal effect of a separate CW laser, and we observe cooling directly through a reduction in the soliton microcomb's optical linewidth from 2.2 MHz at ambient temperature to 280 kHz at 84 K. Our experiments highlight the role that microscopic physics plays in the physical operation and applications of emerging integrated nonlinear photonics. Moreover, in our nonlinear system, laser cooling plays a fundamental role in improving measurement precision with soliton microcombs, which we demonstrate by a nearly 10-dB enhancement in signal-to-noise ratio (SNR) $f_{\rm{ceo}}$ photodetection at our effective base temperature.

Figure 1a-c illustrates the concept of our experiments and our key observations of thermal decoherence in soliton microcombs. Our photonic device is a planar, waveguide-coupled SiN ring resonator; see Fig. 1a. By pumping mode $m$ of the resonator at frequency $\nu_m\approx c/1546 \, \rm{nm}$ with a tunable, amplified CW laser, we generate a few-cycle-duration soliton with repetition frequency $f_{\rm{rep}}\approx$ 1.01 THz, which corresponds to the roundtrip time. Our experiments and analysis show that $f_{\rm{rep}}$ of the emitted soliton pulsetrain is influenced by thermal noise through the thermorefractive effect. Here the refractive index $n(T)$ is a thermodynamic variable subject to Eq. \ref{eq:FD}\cite{gorodetsky2004fundamental, matsko2007whispering}, and this links soliton fluctuations to the resonator temperature $T$. To understand this relationship, we consider the thermorefractive frequency fluctuations of $\nu_m$, using the Langevin equation
\begin{equation}
\frac{d(\delta\nu_m)}{dt}=-\Gamma_T \, \delta\nu_{m}+\zeta(t),
\label{eq:Langevin} \end{equation}
where $\Gamma_T$ is the thermal relaxation rate, $\zeta(t)$ is a stochastic source defined by its autocorrelation, $\left<\zeta(t)\zeta(t+\tau)\right>=\eta_{m}^2\frac{2\Gamma_{T} \, k_B T^2}{\rho \, C \, V}\delta(\tau)$, $\eta_m=\frac{d\nu_m}{dT}$ is the thermal tuning of the cavity modes, and $\delta (\tau)$ is the Dirac delta function\cite{sun2017squeezing}. Defining the thermal decoherence time, $1/\Gamma_T$, leads to a time-domain interpretation of how a soliton is affected by thermal noise. Solitons are periodically outcoupled from the resonator at period $t_{\rm{rep}}=1/f_{\rm{rep}}$ and with stochastic noise $\delta t_{\rm{rep}}$. Hence, repeated measurements of $t_{\rm{rep}}$ at time intervals $\tau$ become uncorrelated for $\tau >>1/\Gamma_{T}$, yielding a distribution of measurement results (Fig. 1a) according to $\left< \delta f_{\rm{rep}}^2 \right> = \eta_{\rm{rep}}^2\frac{k_B T^2}{\rho \, C \, V}$, where $\eta_{\rm{rep}}$ is the coupling coefficient for thermal noise and $f_{\rm{rep}}$; we have measured $\eta_{\rm{rep}}= 32.4$ MHz/K through the characterization reported in Fig. 2.  

To explore thermal decoherence, we use $f$--$2f$ self-referencing to detect $f_{\rm{ceo}}$ and perform systematic experiments to characterize it. The design of our photonic chip enables soliton microcombs with an octave-bandwidth spectrum, including two dispersive-wave peaks (Fig. 1b), and systematic variation of $f_{\rm{ceo}}$ for an electronically detectable microwave frequency. With soliton microcombs, the mode-frequency relationship of Eq. \ref{eq:COMB} yields the coherent signal frequency relationship $f_{\rm{ceo}}=\nu_{\rm{p}}-mf_{\rm{rep}} \inlineeqnum\label{eq:FCEO}$, where $\nu_{\rm{p}}$ is the pump-laser frequency. As a result, thermal fluctuations are phase-coherently multiplied according to $\delta f_{\rm{ceo}}=-m \times \delta f_{\rm{rep}}$; see the schematic in Fig. 1b. This process is central to the concept of any frequency comb through the $n \, f_{\rm{rep}}$ factor in Eq. \ref{eq:COMB} \cite{Hansch2006nobel}. Indeed, the red data in Fig. 1c shows the $f_{\rm{ceo}}$ optical lineshape, which is the $f$--$2f$ optical-heterodyne beatnote. The measured full-width at half-maximum (FWHM) is $\kappa_{\rm{ceo}} = 2.2$ MHz, and this linewidth is substantially larger than the contribution from the pump laser, which is shown by the gray trace in Fig. 1c. To understand this, we model the thermal-noise-limited $f_{\rm{ceo}}$ lineshape (red line in Fig. 1c) based on the relationship between the power-spectral density of $f_{\rm{ceo}}$ fluctuations, which we label $S_{ff}$, and $\left<\delta f_{\rm{ceo}}^2 \right> = m^2\left< \delta f_{\rm{rep}}^2 \right>$. Therefore, we expect the lineshape has a gaussian form with $\kappa_{\rm{ceo}}=\sqrt{8 \ln(2) \, m^2 \, \eta_{\rm{rep}}^2 \, \frac{k_B T^2}{\rho \, C \, V}} \inlineeqnum\label{eq:kappaceo}$, which evaluates to 2.3 MHz at $T=300$ K; see Methods. This observation demonstrates thermal decoherence through a broadened $f_{\rm{ceo}}$ signal, which forms the link between the terahertz and optical-frequency domains. 

\begin{figure*}  \centering 
\includegraphics[width=\linewidth]{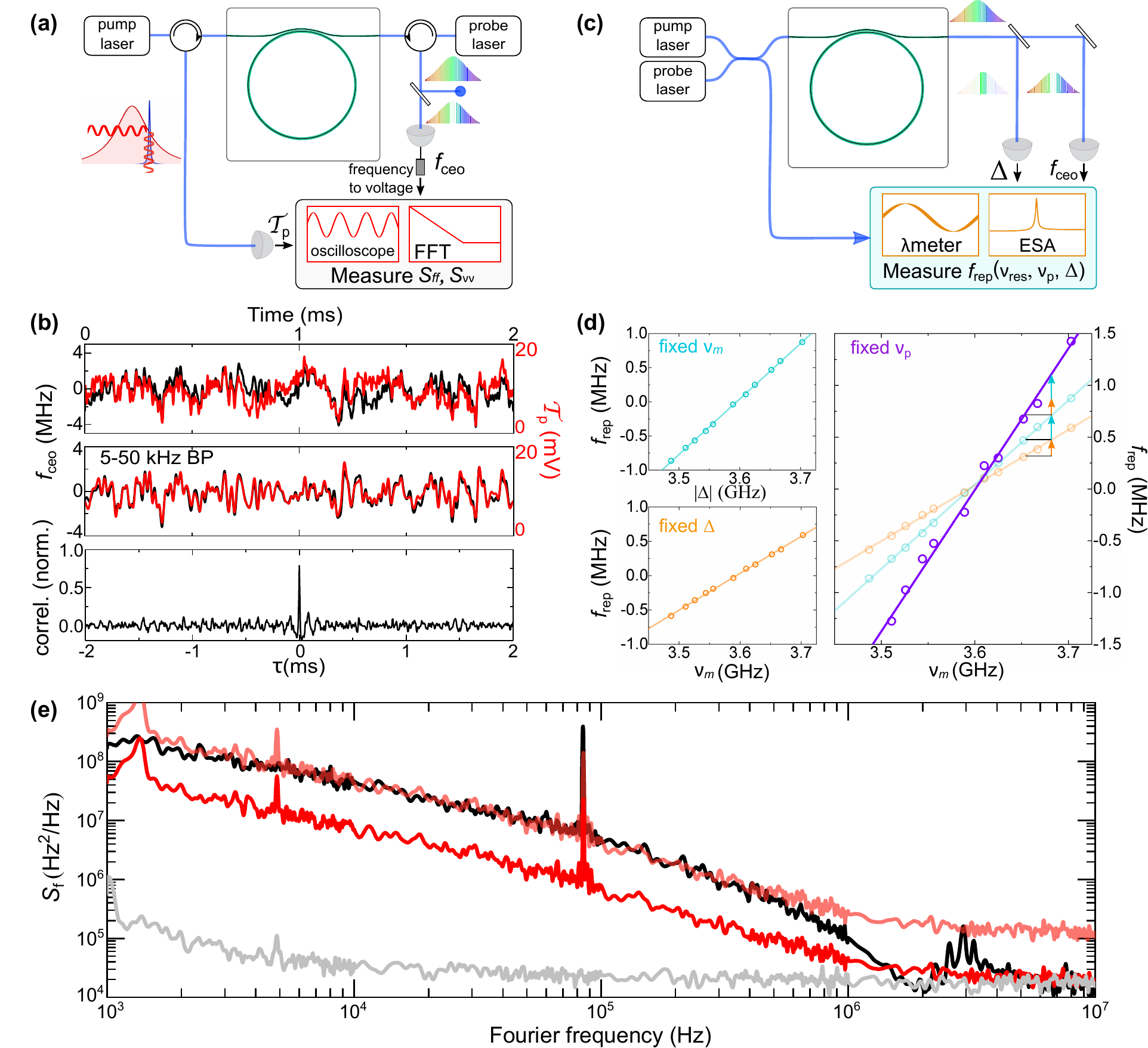}
\caption{Observation and calibration of soliton thermal-noise coupling. (a) Schematic for synchronous measurement of $\nu_{m}$ and $f_{\rm{ceo}}$ with an external probe laser. (b) Top and middle panels: Realtime measurements of probe-laser transmission and $f_{\rm{ceo}}$. A digital bandpass filter (BP) from 5 to 50 kHz is applied in the middle panel. Bottom panel: Normalized correlation of the probe transmission and $f_{\rm{ceo}}$. The peak at $\tau=0$ indicates strong thermal-noise correlations. (c) Thermal-coupling coefficient calibration. We measure $f_{\rm{rep}}$ versus $\Delta$ and $\nu_{m}$ for use in Eq. \ref{etarep}. (d) Coupling coefficients for $f_{\rm{rep}}$ versus $\Delta$ (green), $\nu_{m}$ (orange), and temperature (in units of $\nu_{m}$) for fixed $\nu_{\rm{p}}$ (purple). Here we shifted $\nu_m$ to the displayed range. (e) Power-spectral density measurements of $f_{\rm{ceo}}$ ($S_{ff}$, black trace), $\nu_m$ ($S_{\nu\nu}$, red trace), inferred $S_{ff}$ from thermal noise and coupling coefficient (pale red trace), and measurement noise floor (gray trace).} \label{fig2}
\end{figure*} 

The coefficient $\eta_{\rm{rep}}\equiv df_{\rm{rep}}/dT$ that describes the soliton's thermal-noise coupling is critical to understanding our system and measurements. Importantly, $\eta_{\rm{rep}}$ has not been considered in terms of thermal noise with soliton microcombs, therefore we carry out a detailed analysis and two sets of thermal-noise-calibration experiments. In essence, we seek to define the coupling $S_{ff}(\omega) = m^2 \, \eta_{\rm{rep}}^2 \, S_{TT}(\omega) = m^2 \, \eta_{\rm{rep}}^2 S_{\nu \nu}(\omega)/\eta_{m}^2$, where $\omega$ is the angular Fourier frequency, and $S_{TT}$ and $S_{\nu\nu}$ are the power-spectral densities of resonator thermal noise and resonator-mode frequency noise, respectively. Clearly $S_{TT}=S_{\nu\nu}/\eta_m^2$, where $\eta_m=d\nu_{m}/dT \approx2$ GHz/K is typical of SiN\cite{xue2016thermal}. To understand $\eta_{\rm{rep}}$, we assume the repetition frequency $f_{\rm{rep}}=D_{1}+\frac{\Omega(\Delta)}{2 \pi}\frac{D_2}{D_1}$, where $D_1$ is the microresonator free-spectral-range (FSR), $D_2$ is the quadratic term of the group-velocity dispersion (GVD) expanded around $\nu_m$, and $\Omega(\Delta)$ is a frequency shift in the soliton carrier wave that depends on the pump-resonator detuning, $\Delta=\nu_{\rm{p}}-\nu_m$\cite{yi2017single}. However, thermal noise is coupled to the soliton through $\nu_m$ and $\Delta$, therefore we decompose $\eta_{\rm{rep}}$ as 
\begin{equation}
\eta_{\rm{rep}}=\frac{df_{\rm{rep}}}{d\nu_m}\frac{d\nu_m}{dT}=\left(\frac{\partial f_{\rm{rep}}}{\partial \nu_m}-\frac{\partial f_{\rm{rep}}}{\partial \Delta}\right)\eta_{m}, \label{etarep}
\end{equation}
and we assess that $\partial f_{\rm{rep}}/\partial \nu_m$ captures thermal changes in $D_{\rm{1}}$, while $\partial f_{\rm{rep}}/\partial \Delta$ quantifies the contribution from $\Omega$. In the first set of $\eta_{\rm{rep}}$ measurements, we characterize the thermal-noise correlation of $\nu_m$ and $f_{\rm{rep}}$ with a soliton circulating in the resonator by use of the configuration described in Fig. 2a and Methods. To measure the fluctuations of $\nu_m$, we use a probe laser detuned to one-half linewidth higher frequency than the resonator mode $\nu_{m+1}$ and directed reverse to the pump laser. We photodetect the probe laser intensity noise that is converted from $\nu_m$ noise. In addition, we downconvert the $f_{\rm{ceo}}$ microwave signal to a baseband frequency and record it by use of an analog frequency-to-voltage circuit. With an oscilloscope, we simultaneously detect these fluctuating signals, which are presented in Fig. 2b after processing with two digital-filter functions. Their correspondence is clear, resulting in a strong peak in the cross-correlation (lower panel of Fig. 2b) that reveals the coherent driving of soliton temporal fluctuations by thermal noise.

To quantify thermal-noise coupling to the soliton, we characterize $\partial f_{\rm{rep}}/\partial \nu_m$ and $\partial f_{\rm{rep}}/\partial \Delta$ with the system shown in Fig. 2c. With a soliton circulating in the microresonator, we simultaneously monitor $f_{\rm{rep}}$, $\nu_m$, and $\Delta$. We detect $f_{\rm{rep}}$ by electro-optic modulation\cite{drake2018kerr}; we use the probe laser to monitor $\nu_m$, measuring with a wavemeter, and we determine $\Delta$ from the frequency difference of the probe laser and the pump laser. In our second set of measurements (Fig. 2d), we explore Eq. \ref{etarep} as the chip temperature is varied with a thermoelectric cooler. This calibration procedure yields the results: $\partial f_{\rm{rep}}/\partial \nu_m=5.2$ MHz/GHz; $\partial f_{\rm{rep}}/\partial \Delta = -8.5$ MHz/GHz; and $\eta_{\rm{rep}}=$32.4 MHz/K. Interestingly, despite having different physical origins, in our present experiments $\partial f_{\rm{rep}}/\partial \nu_m$ and $\partial f_{\rm{rep}}/\partial \Delta$ contribute to $\eta_{\rm{rep}}$ with similar couplings and positive signs due to the exact relationship between $\nu_{m}$ and $\Delta$. 

With an understanding of thermal-noise coupling to the soliton, we present experiments and analysis of $S_{ff}$ and $S_{\nu\nu}$ that provide a quantitative picture of how thermal noise impacts the soliton microcomb. We obtain these power-spectral densities by Fourier transform of the analog time-domain $f_{\rm{ceo}}$ and probe-laser intensity noise signals, respectively. Conversely, we use $\eta_m^2$ to characterize $S_{TT}$. Fig. 2e shows $S_{ff}$ (black trace) and $S_{\nu\nu}$ (red trace), and we focus on the frequency range from 1 kHz to 1 MHz that is dominated by thermal noise. These signals are readily measurable in our system. Indeed, our measurement noise floor is 40 dB below these thermal-noise-limited signals (gray trace in Fig. 2e), which indicates the substantial role that thermal noise plays and the advantage of measuring with an $f$--$2f$ interferometer. Although we have not developed a detailed thermal conduction model of the SiN resonator\cite{Huang2019TRnoise}, the frequency dependence of $S_{ff}$ and $S_{\nu\nu}$ are consistent with the 1-2 $\mu$s measured and predicted principal thermal time constant of our resonator. At lower frequencies technical noise of the pump-laser frequency plays a more important role. Using our measurement of $\eta_{\rm{rep}}$, we directly compare the inferred spectrum ($m^2 \, \eta_{\rm{rep}}^2 \, S_{TT}(\omega)$) of $f_{\rm{ceo}}$ (pale red trace in Fig. 2e) with our $f$--$2f$ measurement of $f_{\rm{ceo}}$. The agreement of these data across three decades in Fourier frequency is confirmed by the similar response versus frequency, and it shows the accuracy of our $\eta_{\rm{rep}}$ calibration.

\begin{figure}[h] \centering 
\includegraphics[width=\linewidth]{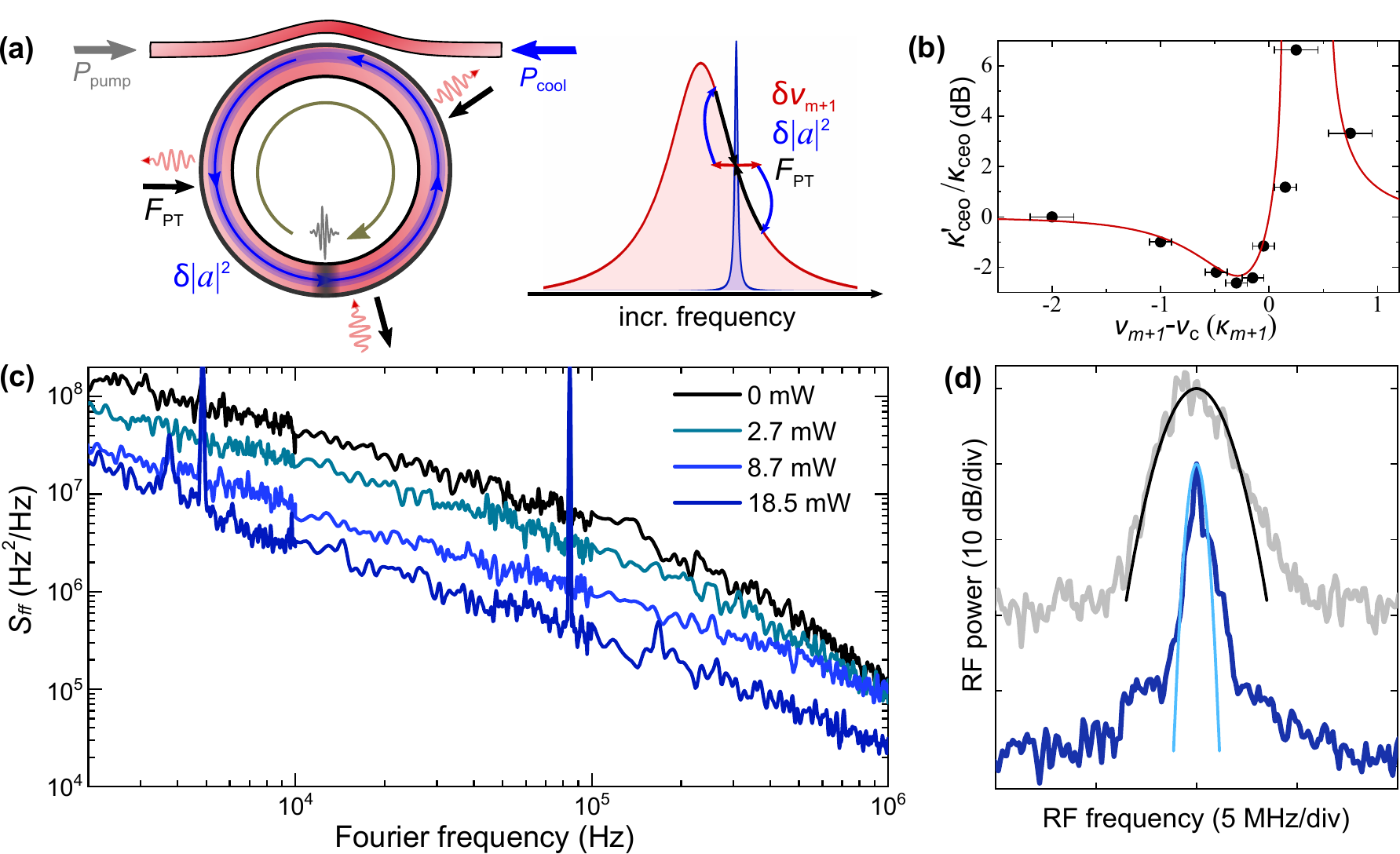}
\caption{Soliton laser cooling. (a) Conceptual illustration of the photothermal forcing, $F_{\rm{PT}}$, used to dynamically maintain $\Delta_{\rm{c}}$. We tune the cooling laser approximately one-half linewidth higher frequency than $\nu_{m+1}$. Thermorefractive noise causes a fluctuation in the intraresonator energy, $\left|{a}\right|^2$, which the cooling laser passively counters. (b) Measured $f_{\rm{ceo}}$ linewidth versus $\Delta_{\rm{c}}$ for $P_{\rm{c}} =$ 5 mW. The linewidth model shown by the red line is based on Eq. \ref{eq:kappaceo} and the photothermal cooling effect $\Gamma_{T}'/\Gamma_{T}$. (c) $S_{ff}(\omega)$ for $P_{\rm{c}} =$ 0, 2.7, 8.7, and 18.5 mW. (d) Measured $f_{\rm{ceo}}$ power spectrum without cooling (black) and for $P_{\rm{c}} =$ 18.5 mW (blue). The data are compared to the model from Eq. \ref{eq:kappaceo}.}\label{fig3}
\end{figure}

In Fig. 3, we introduce laser cooling of the soliton that circulates in our microresonator, achieved by way of a parametric coupling with an auxiliary coherent laser field. For all the experiments reported here, the pump laser is adjusted to maintain soliton stability and $f_{\rm{ceo}}\approx 1$ GHz. The cooling laser gives rise to a dynamic photothermal forcing\cite{metzger2004cavity,barton2012photothermal}, which counteracts soliton thermal noise for an appropriate setting of the cooling laser ($\nu_{\rm{c}}$) frequency detuning $\Delta_{\rm{c}}=\nu_{m+1}-\nu_{\rm{c}}$. (We have observed that laser cooling is also effective by use of other resonator mode numbers $m$, even $m+90$.) Fig. 3a shows a schematic of our experiments and an illustration of how resonator-mode thermal fluctuations necessitate intraresonator power fluctuations. In the resonator, the cooling laser light counterpropagates with respect to the pump and the soliton; the laser is blue-detuned with $\Delta_{\rm{c}} \approx \kappa_{m+1}/2$, where $\kappa_m$ is the resonator-mode linewidth. Photothermal forcing dynamically maintains $\Delta_{\rm{c}}$, countering changes in the intraresonator power that arise primarily from thermal noise. The result is not only a reduction in $S_{ff}$, but a reduction in the $f_{\rm{ceo}}$ linewidth and interestingly an absolute increase in the SNR of the soliton microcomb modal lineshape.

To understand the physical mechanisms involved, we explore how resonator-mode-frequency noise depends on cooling-laser absorption for a photothermal force ($F_{\rm{PT}}=\alpha_{T} \, |a|^2$) and the thermorefractive Langevin force ($\zeta(t)$) according to
\begin{equation}\label{LEcool}
    \frac{d(\delta\nu_{m})}{dt}=-\Gamma_{T} \,\delta \nu_{m}+\alpha_{T} |a|^2+\zeta(t),
\end{equation}
where $\alpha_{T}$ is the thermo-optic heating coefficient, and the cooling-laser intraresonator energy $|a|^2=\kappa_e \, P_{\rm{c}} / \left(\Delta_{\rm{c}}^2 + (\kappa_{m+1}/2)^2\right)$ depends on the external coupling rate $\kappa_e$, the cooling laser power $P_{\rm{c}}$ that is coupled into the on-chip access waveguide, and the resonator-mode linewidth $\kappa_{m+1}$\cite{metzger2004cavity}. Therefore, laser cooling modifies the thermal-relaxation rate of a soliton microcomb to $\Gamma_T'=\Gamma_T+\frac{2\,|\alpha_{T}|\kappa_{e}\Delta_{\rm{c}}}{(\Delta_{\rm{c}}^2+(\kappa_{m+1}/2)^2)^2} P_{\rm{c}} \inlineeqnum\label{eq:gp}$, either increasing or decreasing it according to the sign of $\Delta_{\rm{c}}$. According to how photothermal laser cooling modifies $\Gamma_T'$, the impact of thermal noise on $S_{ff}$ should be reduced by the ratio $\Gamma_T'/\Gamma_T$\cite{metzger2004cavity,sun2017squeezing,restrepo2011classical}. 

We have discovered that the practical effect of soliton laser cooling is $f_{\rm{ceo}}$ frequency-noise reduction within the resonator thermal bandwidth and a correspondingly reduced effective soliton temperature. Therefore, according to our understanding from Eq. \ref{eq:kappaceo}, we expect a reduction in the optical linewidth of $f_{\rm{ceo}}$, which we denote as $\kappa_{\rm{ceo}}'$ for observations with laser cooling applied and $\kappa_{\rm{ceo}}$ without. We directly measure the ratio $\kappa_{\rm{ceo}}'/\kappa_{\rm{ceo}}$ (Fig. 3b) for several settings of $\Delta_{\rm{c}}$ with fixed $P_{c}= 5$ mW, and we observe both a clear reduction and an increase in the $f_{\rm{ceo}}$ linewidth, depending on $\Delta_{\rm{c}}$. Furthermore, the Langevin model well-captures this behavior as shown by the red line in Fig. 3b in which we predict the $\Gamma_T'/\Gamma_T$ reduction in $S_{ff}$ and the corresponding linewidth based on Eq. \ref{eq:kappaceo} and Methods. This demonstrates both the utility of our soliton-laser-cooling technique and its connection to external dynamic control of the soliton microcomb.

We explore the $P_{\rm{c}}$-dependence of soliton laser cooling on $S_{ff}$, searching for the conditions that yield the lowest frequency-noise spectrum (Fig. 3c) and the narrowest $f_{\rm{ceo}}$ linewidth (Fig. 3d). In our experiments, for each setting of $P_{\rm{c}}$ we adjust $\Delta_{\rm{c}}$ to maximize laser cooling. With $P_{\rm{c}}=18.5$ mW power coupled onto the chip, we achieve almost 20 dB of $S_{ff}$ noise reduction across more than three decades in Fourier frequency. This broadband behavior is consistent with $\Gamma_T$ and the relatively low-noise properties of our cooling laser. At still higher settings of $P_{\rm{c}}$, we do observe some technical complexities of soliton laser cooling. Both absolute intensity and frequency noise of the cooling laser hinder the cooling efficiency; this behavior has been reported extensively in laser cooling mechanical systems\cite{kippenberg2013phase}. A second and more unique effect of our soliton microcomb system is initiation of parametric oscillation for $P_{\rm{c}} \geq 20$ mW. Therefore some of the cooling power is lost to signal and idler generation. Even constrained to $P_{\rm{c}}\leq 18.5$ mW, soliton laser cooling affects a profound change in the optical lineshape of $f_{\rm{ceo}}$; see Fig. 3d. Here, the uncooled gray trace reflects an $f_{\rm{ceo}}$ linewidth of 2.3 MHz, according to Eq. \ref{eq:FCEO}, whereas the laser-cooled soliton exhibits a substantially reduced linewidth of 280 kHz and surprisingly the SNR of the $f_{\rm{ceo}}$ lineshape is increased by 10 dB. Indeed, laser cooling does not modify the average optical power of the soliton, therefore it increases the peak $f$--$2f$ photocurrent signal. Such an SNR improvement has important implications for soliton microcomb applications, specifically in higher precision digitization that reduces the $f_{\rm{ceo}}$ signal detection error rate by a factor $\propto \rm{erfc}(10^{SNR/20}/\sqrt{2})$ \cite{Sinclair2015compact}. 

\begin{figure}[h!] \centering
    \includegraphics[width=\linewidth]{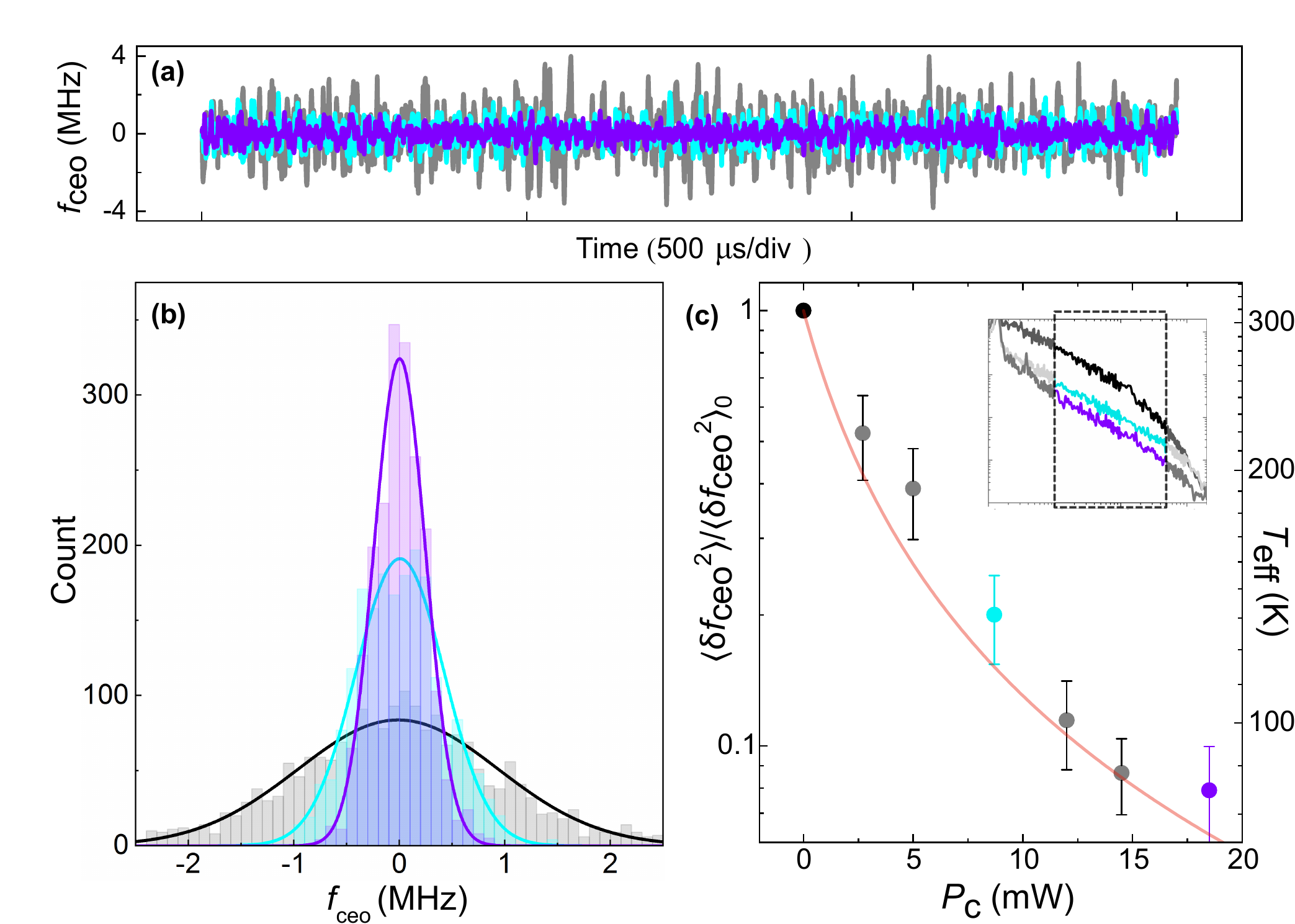}
    \caption{Universal thermal fluctuations. (a) Realtime measurement of $f_{\rm{ceo}}(t)$ for no cooling laser (gray) and $P_{\rm{c}}=$ 8.7 mW (cyan) and 18.5 mW (violet). (b) Direct measurement of the $f_{\rm{ceo}}(t)$ distribution, constructed by binning the data in (a). From these data, we define the effective temperature, $T_{\rm{eff}}$, using its fundamental connection to $\left\langle \delta (f_{\rm{ceo}})^2 \right\rangle$. (c) $T_{\rm{eff}}$ versus $P_{\rm{c}}$. The data from (a) and (b) are indicated by the black, cyan, and violet points. The expected $T_{\rm{eff}}$, based on $\Gamma_T'/\Gamma_T$ and Eq. \ref{eq:teff}, is shown in pale red.}
    \label{fig4}
\end{figure}

Viewing soliton thermal noise and laser cooling through $S_{ff}$ as above necessitates a detailed materials understanding of the resonator that holds the soliton particle. With the complementary set of experiments presented in Fig. 4, we reveal the universal nature of thermal fluctuations traced back to Eq. \ref{eq:FD} by exploring $\left\langle \delta f_{\rm{ceo}}^2 \right\rangle$. Such an analysis makes thermal-noise-limited soliton characterization straightforward through a definition of temperature. Specifically, the connection between an ensemble average of $f_{\rm{ceo}}$ fluctuations and Eq. \ref{eq:FD} enables us to define an effective temperature ($T_{\rm{eff}}$) associated with laser cooling our soliton, according to
\begin{equation}
    T_{\rm{eff}}=\sqrt{\frac{\left< \delta \textit{f }_{\rm{ceo}}^2 \right> \rho \, C \, V}{m^2 \,\eta_{\rm{rep}}^2 \, k_{\rm{B}}}}=T_{\rm{0}}\sqrt{\frac{\left< \delta \textit{f }_{\rm{ceo}}^2 \right>}{\left< \delta \textit{f }_{\rm{ceo}}^2 \right>_{\rm{0}}}}, \label{eq:teff}
\end{equation}
where $T_0$ is the ambient temperature and $\left< \delta {f_{\rm{ceo}}}^2 \right>_{\rm{0}}$ is the variance in $f_{\rm{ceo}}$ without laser cooling. Hence, a real-time record of $f_{\rm{ceo}}$ contains the full set of information to understand soliton thermodynamics, and such measurements are important for active-feedback cooling protocols\cite{wilson2015measurement}.  

Using the frequency-to-voltage circuit and the digital filter as described previously in Fig. 2c, we measure $f_{\rm{ceo}}$ fluctuations in real time and with high sensitivity; see Fig. 4a. We verified that our digital filter, which reduces low Fourier frequency contributions from pump-laser noise, does not influence our conclusions regarding $T_{\rm{eff}}$. Without applying laser cooling we record the gray $f_{\rm{ceo}}(t)$ trace that is characterized by a standard deviation of 0.95 MHz. By activating the laser cooling at moderate (maximum) settings of $P_{\rm{c}}$, we obtain the cyan (violet) $f_{\rm{ceo}}(t)$ trace that shows a decreased standard deviation. We present all three traces as histograms in Fig. 4b, which makes clear that soliton laser cooling reduces $f_{\rm{ceo}}$ noise and increases our measurement precision. We use gaussian fitting to determine $\left< \delta f_{\rm{ceo}}^2 \right>$ and hence $T_{\rm{eff}}$ from the histograms. In our experiments, we vary $P_{\rm{c}}$ and record the variance of $f_{\rm{ceo}}$ normalized to the case of $P_{\rm{c}}=0$; see Fig. 4c. We expect a reduction in $T_{\rm{eff}}$ with increasing $P_{\rm{c}}$, according to Eq. \ref{eq:teff} and the Langevin laser-cooling model for $\Gamma_{T}/\Gamma_{T}'$ that we show with the red line in Fig. 4c. We assess that the laser-cooled soliton effective temperature can reach the ${T_{\rm{eff}}}\approx$ 84 K, which might otherwise require immersion of our entire photonic chip in liquid nitrogen. Moreover, straightforward technical improvements in our system-- such as laser cooling with a resonator mode that experiences normal GVD and with a lower-noise laser-- should enable experiments to reach $T_{\rm{eff}}<10$ K. 

In summary, we have discovered thermal decoherence in Kerr-soliton frequency combs. Our measurements highlight strong thermal-noise correlations between a soliton and the photonic-chip resonator that holds it. The result of thermal decoherence is a frequency-broadened lineshape of the soliton microcomb modes, which imposes a fundamental reduction in measurement precision. However, we have also introduced soliton laser cooling through a passive microresonator photothermal forcing. We reduce the linewidth of a laser-cooled soliton microcomb by nearly a factor of ten, and we observe a commensurate SNR increase. Moreover, we have shown how to characterize a soliton microcomb by an effective temperature, which fully evaluates the coherence of the soliton's electromagnetic field. Integrated-photonics devices, especially those that utilize nonlinearity, are expected to offer scalable solutions to application and measurement problems; Kerr frequency combs are a key example. Still, as our work demonstrates, future innovation will depend on understanding the role that microscopic physics and its control plays in nanophotonic devices.

We thank Kartik Srinivasan for fabricating the SiN microresonators, Su-Peng Yu for creating the mode simulation in Fig. 1, Daryl Spencer for experimental assistance, and Srico, Inc. for use of the PPLN waveguide device. This research is supported by the Defense Advanced Research Projects Agency DODOS program, AFOSR (FA9550-16-1-0016), NRC, and NIST. This work is not subject to copyright in the United States.

\begin{methods}

\section*{Soliton generation and self referencing.} 
We generate Kerr solitons, using the fast-sweeping method\cite{briles2018interlocking, stone2018thermal}. A pump laser (New Focus Velocity) is modulated in the single-sideband, suppressed-carrier configuration by a voltage-controlled oscillator with 10--20 GHz output frequency. The pump-laser frequency is positioned on the blue side of $\nu_m$ and swept red by 10 GHz to a final, red-detuned optical frequency. The frequency sweep initiates modulation instability in the microresonator and subsequently induces its condensation into a Kerr soliton. Optimizing the sweep speed mitigates thermal transients that arise from sudden changes in intracavity power; we find that a $\approx$100 ns sweep time is appropriate for our system.

Our chip contains dozens of photonic circuits to vary the dispersive-wave peak wavelengths and the resonator's mode structure that controls $f_{\rm{ceo}}$. After guiding the soliton microcomb off-chip, we separate the short- and long-wavelength spectral components for $f$--$2f$. The modes near 1965 nm are amplified with thulium-doped fiber\cite{li2013thulium}, frequency-doubled in a periodically poled lithium-niobate (PPLN) waveguide, and recombined with the 982.5-nm comb lines on an avalanche photodiode. Our experiments demonstrate $f$--$2f$ self-referencing of a microcomb without the aid of external broadening\cite{lamb2018optical} or auxiliary lasers\cite{briles2018interlocking}.

Our PPLN waveguide (Srico) is $\approx$3 cm long and temperature-controlled with off-chip heaters. The temperature is adjusted to optimize phase matching, for an on-chip efficiency of 30 \%/W. The frequency-doubled comb light and short-wavelength dispersive wave are combined in a 50\% coupler and collectively amplified to $\approx$10 $\mu$W in a semiconductor optical amplifier. 

\section*{Theory of thermal $f_{\rm{ceo}}$ linewidth.} A relationship exists between $\kappa_{\rm{ceo}}$, $S_{ff}$, and $T_{\rm{eff}}$, which we analyze using Eq. \ref{eq:FD}. We write the FWHM linewidth of an oscillator as $\kappa=\sqrt{8 \,\ln \,(2)\textit{A}}$, where $A$ is the power spectral density of frequency fluctuations integrated up to the so-called beta line\cite{di2010simple}, defined by $S_{\beta\beta}(\omega)=\omega\times 4\rm{ln}(2)/\pi^3$. To predict $\kappa_{\rm{ceo}}$ from the system temperature, we first observe that for $S_{ff}$ at room-temperature, $f_{ceo}$ thermal noise is mostly above $S_{\beta\beta}$. Hence, we connect $T$ and $\kappa_{\rm{ceo}}$ according to
\begin{equation}
    \frac{1}{2\pi}\int S_{ff} \, d\omega=m^2\eta_{\rm{rep}}^2\frac{k_{\rm{B}}T^2}{\rho CV}\approx \frac{\kappa_{\rm{ceo}}^2}{8\,\ln(2)},
\end{equation}
from which we predict a thermal-noise-limited $\kappa_{\rm{ceo}}=\sqrt{8\rm{ln}(2)\times \textit{m}^2\eta_{\rm{rep}}^2\frac{\textit{k}_{\rm{B}}\textit{T}^2}{\rho \textit{CV}}}$. Using $\eta_{\rm{rep}}$ = 32.4 MHz/K (see below), $\rho$ = 2600 kg/m$^3$, $C$ = 650 J/kg-K, and $V$ = 3.0 $\times$ 10$^{-17}$ m$^3$, we calculate $\kappa_{\rm{ceo}}$ = 2.3 MHz, in agreement with the experimental value of 2.2 MHz. A theoretical $f_{\rm{ceo}}$ power spectrum based on this model is shown in Fig. 1c. 

In Fig. 3d, we apply this model to the $f_{\rm{ceo}}$ linewidth reduction from laser cooling. For significant cooling, a non-negligible fraction of $S_{ff}$ is reduced below $S_{\beta \beta}$, resulting in a narrower $f_{\rm{ceo}}$ spectrum than predicted by the model. While the exact discrepancy depends on the form of $S_{ff}$, in our system we experimentally achieve $\kappa'_{\rm{ceo}}=$ 280 kHz, compared to $\approx$ 640 kHz from the model. We have verified the linewidth reduction through a numerical study of $S_{ff}$. 

\subsection{Cross correlation measurement.} We operate the probe laser blue-detuned from $\nu_{m+1}$ by slightly less than a half linewidth. At this detuning, frequency fluctuations in either the probe laser or microresonator are converted in a calibrated fashion to the probe-laser intensity. The probe-laser power is set to $<$1 mW of chip-coupled power to prevent noise coupling between the probe and microresonator. The intrinsic intensity noise and phase noise of the probe laser are measured separately and are verified to not significantly contribute to the signal. Therefore, we conclude that the probe transmission is a reliable measurement of $\nu_{m+1}$ frequency fluctuations. 

\subsection{Calculation of $f_{\rm{rep}}$ tuning coefficients.} The frequency-comb repetition rate is determined by the soliton group velocity and microresonator path length as
\begin{equation}
f_{\rm{rep}}=c/n_{\rm{g}}L, \label{eq:frep}
\end{equation}
where $c$ is the speed of light in vacuum and $n_{\rm{g}}$ is the group index. Neglecting changes in $L$, the tuning of $f_{\rm{rep}}$ with temperature $T$ is given by
\begin{equation}
\frac{d f_{\rm{rep}}}{dT}=\frac{f_{\rm{rep}}}{n_{\rm{g}}}\frac{d n_{\rm{g}}}{d T}. \label{eq:nurep decomp}
\end{equation}
Due to resonator dispersion, $n_{\rm{g}}$ is a function of the soliton carrier-wave frequency, $\omega_{\rm{s}}$, which also has a temperature dependence. This separate dependence stems from the detuning $\Delta$ between the pump laser and the cavity resonance, the latter being temperature sensitive. The shift in $\omega_{\rm{s}}$ with $\Delta$ is known as the soliton self-frequency shift (SSFS) and has been extensively explored in the literature\cite{karpov2016raman,yi2017single}. The derivative on the right side of Eq. \ref{eq:nurep decomp} must therefore be separated into its constituent parts as 
\begin{equation}
\frac{dn_{\rm{g}}}{dT}=\frac{\partial n_{\rm{g}}}{\partial T} + \frac{\partial n_{\rm{g}}}{\partial \omega_{\rm{s}}} \frac{\partial \omega_{\rm{s}}}{\partial \Delta} \frac{\partial \Delta}{\partial T}.
\end{equation}
The first term on the right side corresponds to a thermal shift of the entire $n_{\rm{g}}(\omega_{\rm{s}})$ curve, while the second term describes temperature-induced movement along the curve. In past analyses of the soliton thermal dynamics\cite{xue2016thermal}, only the first term has been considered. This leads to the prediction
\begin{equation}
\frac{df_{\rm{rep}}}{d\nu_{m}}\approx\frac{f_{\rm{rep}}}{\nu_{m}}.
\end{equation}
For a cavity resonance frequency of $\approx$194 THz, this yields a tuning coefficient of $\approx$5.2 MHz/GHz. 

We use this descritpion of Kerr-soliton dynamics to understand the measurements presented in the main text. Both $\nu_{m}$ and $\Delta$ are sensitive to changes in either $T$ or $\nu_{\rm{p}}$. Hence, we study the actions of $\nu_{m}$, $\Delta$, and $f_{\rm{rep}}$ as a system of equations against controlled changes in $T$ and $\nu_{\rm{p}}$. To this end, we measure
\begin{equation}
\frac{df_{\rm{rep}}}{d\Delta_{T}}=\frac{\partial f_{\rm{rep}}}{\partial \Delta} + \frac{\partial f_{\rm{rep}}}{\partial \nu_{m}}=13.5 \, \rm{MHz/GHz},
\end{equation}
\begin{equation}
\frac{df_{\rm{rep}}}{d\Delta_{\rm{\nu}}}=\frac{\partial f_{\rm{rep}}}{\partial \Delta}+\frac{\partial f_{\rm{rep}}}{\partial \nu_{m}}\frac{\partial \nu_{m}}{\partial \nu_{\rm{p}}}\frac{\partial \nu_{\rm{p}}}{\partial \Delta}=-25 \,\rm{MHz/GHz},
\end{equation}
where $\Delta_{T}$ implies a temperature-actuated detuning and $\Delta_{\nu}$ indicates a pump-frequency-actuated detuning. Multiplying $\frac{df_{\rm{rep}}}{d\Delta_{T}}$ by the SiN thermal tuning, $\frac{d\nu_{m}}{dT}$ = 2.4 GHz/K gives $\eta_{\rm{rep}}$ = 32.4 MHz/K. Furthermore, subtracting the above equations cancels out the shared term and leaves 
\begin{equation}
\frac{\partial f_{\rm{rep}}}{\partial \nu_{m}}\left(1-\frac{\partial \nu_{m}}{\partial \nu_{\rm{p}}}\frac{\partial \nu_{\rm{p}}}{\partial \Delta}\right)=38.5 \rm{\,MHz/GHz}.
\end{equation}
Taking the derivative of $\nu_{m}=\nu_{\rm{p}}-\Delta$ and using a measured coefficient of $\partial \nu_{\rm{p}}/\partial \Delta$ = -6.1 GHz/GHz gives $\partial f_{\rm{rep}}/\partial \nu_{m}$ = 5.2 MHz/GHz, which agrees well with the prediction. We thus determine that the coefficients measured in Fig. 2 are consistent with the basic theory outlined above, and that understanding the thermal noise in any nonlinear, microresonator-based frequency conversion process requires consideration of both the temperature-dependent refractive index and detuning-dependent nonlinear processes that are coupled to the temperature.

\subsection{Effective damping rate.} To approximate $\Gamma_T'$ we expand $|a|^2$ around $\delta\nu_{m}=0$, 
\begin{equation}
    |a|^2=\frac{\kappa_{\rm{e}}P_{\rm{c}}}{\Delta_{\rm{c}}'^2+(\frac{\kappa_{m+1}}{2})^2}=\frac{\kappa_{\rm{e}}P_{\rm{c}}}{\Delta_{\rm{c}}^2+(\frac{\kappa_{m+1}}{2})^2}+\frac{2\Delta_{\rm{c}}\kappa_{\rm{e}}P_{\rm{c}}}{\big[\Delta_{\rm{c}}^2+(\frac{\kappa_{m+1}}{2})^2\big]^2}\delta\nu_{m+1}+...
\end{equation}
The zeroth-order term corresponds to a static shift in $\nu_{m+1}$ that does not contribute to the photothermal dynamics. Moreover, higher-order terms become increasingly negligible so long as the $\nu_{m+1}$ frequency jitter is much less than its linewidth ($\delta\nu_{m+1}<<\kappa_{m+1}$). We therefore consider only the first-order term in the Langevin analysis. This substitution immediately yields Eq. \ref{eq:gp}.

In the low-$P_{\rm{c}}$ limit, which we use as a basic theoretical comparison to our results, the $f_{\rm{ceo}}$ noise reduction can be written 
\begin{equation}
    \left< \delta f_{\rm{ceo}}^2 \right> = \frac{\Gamma_{T}}{\Gamma_{T}'}\left< \delta f_{\rm{ceo}}^2 \right>_{\rm{0}}.
\end{equation}
The theory curve in Fig. 4c is derived from this expression with $\Gamma_{T}\approx100$ kHz, which is consistent with measured and predicted resonator thermal time constants.

\end{methods}

\newpage
\bibliographystyle{naturemag}
\bibliography{Thermal}

\end{document}